# A Mechanically Tunable Quantum Dot in a Graphene Break Junction

S. Caneva,[1] M. D. Hermans,[1] M. Lee,[1] A. García-Fuente,[2,3] K. Watanabe,[4] T. Taniguchi,[4] C. Dekker,[1] J. Ferrer,[2,3] H.S.J. van der Zant,[1] and P. Gehring[1,5*]

[1]Kavli Institute of Nanotechnology, Lorentzweg 1, 2628 CJ, Delft, The Netherlands.
[2]Departamento de Física, Universidad de Oviedo, 33007 Oviedo, Spain.
[3]Centro de Investigación en Nanomateriales y Nanotecnología, Universidad de Oviedo – CSIC, 33940 El Entrego, Spain.
[4]Advanced Materials Laboratory, National Institute for Materials Science, 1-1 Namiki, Tsukuba, 305-0044, Japan.
[5]IMEC, Kapeldreef 75, B-3001, Leuven, Belgium.

*Corresponding author. Email: pascal.gehring@imec.be

**Abstract**

Graphene quantum dots (QDs) are intensively studied as platforms for the next generation of quantum electronic devices. Fine tuning of the transport properties in monolayer graphene QDs, in particular with respect to the independent modulation of the tunnel barrier transparencies, remains challenging and is typically addressed using electrostatic gating. We investigate charge transport in back-gated graphene mechanical break junctions and reveal Coulomb blockade physics characteristic of a single, high-quality QD when a nanogap is opened in a graphene constriction. By mechanically controlling the distance across the newly-formed graphene nanogap, we achieve reversible tunability of the tunnel coupling to the drain electrode by five orders of magnitude, while keeping the source-QD tunnel coupling constant. These findings indicate that the tunnel coupling asymmetry can be significantly modulated with a mechanical tuning knob and has important implications for the development of future graphene-based devices, including energy converters and quantum calorimeters.

**Introduction**

The ability to precisely manipulate individual charge carriers is a cornerstone for devices ranging from single electron transistors (SET) to solid state quantum bits (qubits). Graphene exhibits weak spin-orbit and hyperfine interactions, leading to long spin coherence times, and is therefore considered a suitable platform to host qubits (1). Quantum dots (QDs) are at the heart of these applications, with a variety of new structures enabling increasingly more accurate control over the localization, energy and coherence times of the charge carriers (2, 3).

Graphene, however, has two significant limitations: the absence of a bandgap and the occurrence of Klein tunneling, which in practice render the confinement of carriers challenging. The most widely used approaches to produce the required confinement rely on (i) lithographically defining a physical QD in graphene monolayers(4), or (ii) opening a bandgap in bilayer graphene through the application of a vertical electric field, in combination with several local gates to electrostatically confine the carriers (5, 6). Additional improvements to the device quality have recently been achieved with hBN encapsulation to reduce the disorder potential, as well as the use of a graphite back gate to screen impurities in the substrate and to provide a flat support, thereby enhancing the mobility of charge carriers (7). In graphene QD devices, complete current pinch-off has been achieved allowing the study of electron-hole crossover (8), the excitation spectrum (9), spin and valley states (6, 10), and charge relaxation times (11). The control over the transparency of the tunnel barriers is more challenging and is typically addressed by designing QDs connected via long narrow graphene constrictions (11). While the tunnel couplings can be tuned individually, the range over which they can be controlled remains modest.

Here, we report on a new device architecture that provides exquisite control of the tunnel coupling through the controlled rupture of a graphene nano-bowtie in a three-point bending geometry. The device consists of a monolayer graphene mechanical break junction, with a graphite back gate integrated in a van der Waals heterostructure. The atomic thinness of the graphene electrodes reduces the electrostatic screening of an applied gate voltage while an hBN gate dielectric is used as an atomically flat and flexible support that is relatively free of charge traps (12), ensuring high quality graphene/hBN/graphite interfaces. The device has the combined capability of (i) ultrastable mechanical adjustments of the electrode-electrode distance at the nanoscale level and (ii) electrostatic gating. This dual implementation enables a detailed characterization of electronic transport that would not be possible if these tuning parameters were addressed separately. In our design, we present a mechanically tunable monolayer graphene QD formed during breaking of the nanoconstriction at room temperature in air. Low-temperature electronic-transport characterisation reveals high-quality QD electronic properties, where the high gate coupling factor of $\alpha = 0.2$ of the device allows us to address the single electron/hole regime and fill the QD up with N ~80 electron/holes. Furthermore, the device architecture allows mechanical tunability of *both* the tunnel coupling and the capacitive coupling between the QD and the electrodes. Specifically, our methodology enables control over the strength and symmetry of the tunnel couplings to the source and drain leads.

Such a full and reversible manipulation of a graphene QD, in which the degree of confinement of carriers can be controlled both mechanically and electronically, is unique and relevant for applications where tunnelling asymmetry is crucial for device performance (13), such as in quantum calorimetry (14) and in QD energy harvesters (15, 16).

## RESULTS

### Room-Temperature Characterization

Figure 1 (A and B) shows a schematic of the gated graphene break junction device and an optical image of the device prior to bending, respectively. In short, a graphene flake is lithographically patterned into a bowtie shape with a constriction width of ~160 nm. The bowtie is supported on an exfoliated hBN flake (thickness ~30 nm), which acts as a dielectric layer that partially covers a graphite flake (thickness ~30 nm) that acts as the back gate. The van der Waals heterostructure (graphene/hBN/graphite) is assembled on a silicon wafer using an all-dry stamping technique (17) and subsequently transferred on a flexible phosphor bronze substrate coated with insulating polymer layers (PMGI/PI). The flexible substrate allows bending of the device in a three-point bending geometry. A side view of the stack of the final device is shown in Fig. 1C.

The current between the source and drain, $I$, as a function of gate voltage, $V_G$, is initially measured at room temperature in air for a bias voltage, $V$, of 100 mV. Figures 1D-H show the variation of $I$ over a 10 V gate voltage range ($V_G$ from -5 V to 5 V) for different amounts of substrate bending. The unbent substrate (Fig. 1D) displays a minimum in conductance at 0.8 $V_G$, which can be attributed to the charge neutrality point (CNP) of the graphene device. The presence of the CNP at such low $V_G$ attests to the high quality of the sample, where the contribution from substrate doping is minimized by the presence of the graphite/hBN support. For the strained junction (Fig. 1E), the shape of the gate trace is not significantly changed, with the CNP remaining at the same gate voltage position. The slopes around the CNP, however, become steeper, possibly indicating a change in the capacitive coupling to the gate as any wrinkles/folds in the graphene flake are smoothed out during stretching (see Fig. S1). A recent study demonstrated that ripples and corrugations can be present in hBN-supported graphene devices, leading to random strain fluctuations (18). Importantly, these can be reduced by uniaxially straining the device, thereby leading to a charge carrier mobility enhancement. After sufficient strain is applied to the junction, the graphene bowtie breaks and $I$ drops by several orders of magnitude, from µA to nA indicating that a nanogap has formed (Fig. 1F). Upon reversing the direction of bending (i.e. unbending), $I$ starts to rise and a dependence on the back gate voltage is re-established, evidenced by the reappearance of the CNP (Fig. 1G inset). With continued re-approaching of the graphene edges, the initial shape of the CNP is re-established and $I$ recovers to µA levels (Fig. 1H). The opening/closing of the nanogap was performed five times on the same device in air and the electronic behavior was consistent for each cycle (i.e. the $I$-$V_G$ characteristics are unchanged between each cycle) demonstrating that mechanical displacement is not detrimental to the device operation and is reversible. The performance is similar to our previous work on two-terminal graphene break junctions (19), in which electrical contact between the graphene edges is recovered due to the formation of a bilayer overlap region during unbending.

**Coulomb blockade at 4K**

Following the room-temperature characterization described above, we transferred the sample to a bending stage inside a cryostat operated at 4.2 K. The effect of bending height on the conductance of the junction is shown in Figure S1, which plots the last gate voltage traces before opening a nanogap, as indicated by the progressive loss of a current dependence on the gate voltage. Measurements of the zero-bias differential conductance as a function of gate voltage exhibit sharp conductance peaks separated by low conductance regions (Fig. 2A). The separation between the conductance peaks is comparable after each mechanical cycle (i.e. opening and closing of the nanogap), indicating a constant gate coupling strength. The peak heights are strongly reduced in the transport gap around the CNP of graphene (0 V<$V_G$<2 V). Figure 2 (B-D) shows $I$ as a function of bias (±25 mV) and gate voltage (±5 V) at the point where the nanogap is just formed. By sweeping the gate voltage we can tune the carriers from holes (negative $V_G$) to electrons (positive $V_G$). Well away from the CNP, which was at 0.8 $V_G$ for the unbroken graphene sheet, we observe long sequences of regular and closing diamond-shaped regions of suppressed current (white areas). These Coulomb diamonds are characteristic of QD systems in which the energy necessary to add an extra electron to the QD, the addition energy $E_{add}$, exceeds the thermal energy $k_B T$ and in which the tunnel resistances between the QD and the electrodes are much larger than the resistance quantum $h/e^2$. The Coulomb diamonds far away from the CNP are of comparable sizes, with no overlapping features which suggests that a single QD dominates transport through the junction. The diameter $D$ of the QD can be estimated by modelling it as a circular plate capacitor with $D = \sqrt{\frac{4C_G t}{\varepsilon_0 \varepsilon_r \pi}}$, where $C_G$ is the capacitive coupling between the QD and the gate electrode, $\varepsilon_0$ is the vacuum permittivity, and $t = 30$ nm and $\varepsilon_r \approx 4$ are the thickness and the relative dielectric constant of the hBN gate dielectric (20), respectively. The addition energy $E_{add} = \Delta + 2E_c$ is dominated by the charging energy, $E_c$, if we assume that the quantum confinement energy $\Delta$ of the QD is negligible, which puts an lower bound on the effective dot size. Thus $E_{add} \approx 2E_c = \frac{e^2}{C_\Sigma} = \frac{\alpha e^2}{C_G}$, where $C_\Sigma$ is the total capacitance of the QD and $\alpha = \frac{C_G}{C_\Sigma}$ is the lever arm. From the height and width of the diamonds at large positive and negative voltage, we extract $E_{add}$ ~12 meV and $\alpha = 0.2$, which yields $D$ ~60-70 nm. Out tight binding simulations of irregularly-shaped islands ~60 nm in size support the above assumptions, yielding an average $\Delta$ ~ 0.2-0.8 meV, see Fig. S5 in the SI.

We note that the Coulomb diamonds become irregular in shape and spacing near the CNP (Fig. 2C). Such irregular dots around CNP have been observed in transport measurements of graphene constrictions (21, 22) and have been attributed to charge localisation by the formation of charge puddles. The origin of the QD dominating at high/low gate voltages will be discussed below.

## Mechanical tuning of contact transparencies

In Fig. 3 we illustrate the effect of the bending height, $\Delta z$, (which is proportional to the relative in-plane displacement of the two graphene edges) on the transport properties of the junction. We compare the stability diagrams and the conductance peaks for four values of $\Delta z$. Starting from the most 'open' position (Fig. 3, A and E), the first diamonds begin to appear, and the conductance peak amplitudes are relatively small (tenths of nS) with narrow linewidths. The peaks can be fitted using the expression for the classical Coulomb blockade regime ($\Gamma, \Delta \ll k_B T \ll 2E_c$) where $G \sim \cosh^{-2}(e\alpha\delta V_G/2.5 k_B T_e)$, where $\delta V_G = V_G - \delta V_G^{peak}$. This fit yields a $T_e$ of 4.2 K, close to the base temperature of our cryostat, which indicates that the peak broadening is limited by temperature rather than the lifetime of the resonance. These characteristics are evidence of weak coupling of the QD to the reservoirs. Upon closing the junction (Fig. 3, B and F) the sharp-edged diamonds fully close and extra lines parallel to the edges are seen inside the sequential tunnelling regions (i.e., excited states of the QD). Furthermore, the conductance peaks increase in amplitude and in width. Further closing of the gap (Fig. 3, C and G) causes the features in the stability diagram to start blurring, and correspondingly the conductance peaks become broader, such that the tails of adjacent peaks overlap and the baseline conductance acquires a non-zero background. Finally, in Fig. 3, D and H the features are almost completely smeared out although the diamond shapes can still be discerned. The conductance peak amplitudes decrease while their width continue to increase. In this regime the peaks can be fitted with a Breit-Wigner resonance and are characterized by tails that have a slower decay than expected for a thermally broadened peak. This marks the onset of the strong electronic coupling regime.

In the following we investigate the influence of electrode displacement on the capacitive couplings of the QD. To this end, we select a gate range containing one full diamond in the hole regime and monitor it over a range of bending heights (Fig. 4). It is evident that the diamond tilts during closing of the gap, indicating a continuous change in the symmetry of the capacitive couplings. Concurrently, the features become more smeared out, suggesting a change in the contact transparencies (tunnel couplings). Given the strong variation of the electrical properties with electrode displacement, in Fig. 5 we analyse two properties of the system in more detail: (i) the capacitive coupling and (ii) the tunnel coupling of the QD to the source and drain electrodes.

The capacitive couplings of a QD to the source ($C_S$) and to the drain ($C_D$) can be extracted from the positive slope $\beta = \frac{C_G}{C_S}$ and negative slope $\gamma = -\frac{C_G}{C_D + C_G}$ of the Coulomb diamond[23]. In Fig. 5A we plot $C_S$ and $C_D$ extracted from Coulomb diamonds recorded at different bending heights for the same displacement range used in Fig. 4. The gate capacitance is estimated from the addition energy assuming $E_{add} \approx 2E_c$ and using $C_G = \frac{\alpha e^2}{E_{add}}$. Fig. 5A shows that while $C_G$ and $C_S$ remain constant over the displacement range, with values of 3.1 aF and 3.5 aF respectively, $C_D$ is initially (i.e. for larger electrode displacements) lower than $C_S$ but continues to increase from ~4 aF to ~19 aF. This demonstrates that the capacitive coupling of the quantum dot can be tuned by almost a factor of 5 by mechanical displacement of the electrodes.

We now use this data to convert the bending height into an in-plane displacement of the graphene electrodes. To this end we assume, that the drain capacitor which can be tuned mechanically is formed from the drain graphene electrode which partially overlaps with the graphene quantum dot. The overlap area $A = w \cdot d$ is given by the width of the constriction ($w = 160$ nm extracted by atomic force microscopy) and the overlap $d$. By bending the sample $d = d_0 - F\Delta z$ is changed, where $F$ is the attenuation factor of the junction (see Fig. S2). This changes the capacitance by $C_D(d) = \frac{\varepsilon_0 \varepsilon_r w d}{z}$, where $z = 0.335$ nm is the inter-sheet distance for graphene stacks and $\varepsilon_r = 1$. A fit to the data is shown in Fig. 5A which yields the attenuation factor $F$ that we use to convert the bending height into displacement $d$ (see top $x$-axis in Fig. 5A).

In the following we investigate the effect of displacement on the tunnel coupling between the quantum dot and the electrodes. In Fig. 5B the average conductance $\langle G \rangle$ of the Coulomb peak maxima in a given gate voltage window (-4.1 V < $V_G$ < -3.9 V) is plotted over a wider in-plane displacement range of ~6 nm. When increasing the displacement (moving contacts closer together), $\langle G \rangle$ increases to a maximum value of about 40 nS but decreases when $d$ is increased further. Given that $\langle G \rangle$ is a measure of the strength and symmetry of the tunnel couplings to the source and drain leads, the data provides evidence that this symmetry is broken during the closing cycle. More specifically, we attribute the modulation of $\langle G \rangle$ with $d$ to a change of the tunnel coupling $\Gamma_D$ between the QD and the drain electrode while the tunnel coupling $\Gamma_S$ between the QD and the source electrode stays constant. Similar observations of displacement-dependent capacitances and tunnel couplings have also been reported in other material systems, including single-molecule transistors consisting of a $C_{60}$ molecule trapped between two gold electrodes in a mechanical break junction setup (24, 25).

The $\langle G \rangle$ versus $d$ data can be modelled using a Landauer approach (26) in which the maximum conductance can be written as:

$$G(\Gamma_D) = \frac{2e^2}{h}\int_{-\infty}^{\infty}\frac{df(E)}{dE}T(E) = \frac{2e^2}{h}\int_{-\infty}^{\infty}\frac{df(E)}{dE}\frac{\Gamma_S\Gamma_D}{\left(\frac{\Gamma_S}{2}+\frac{\Gamma_D}{2}\right)^2+E^2}dE, \quad (1)$$

where the transmission function $T(E)$ is modelled by a Breit-Wigner resonance, $f(E)$ is the Fermi-Dirac distribution and $\frac{df(E)}{dE} = \frac{1}{4k_BT}\cosh^{-2}\left(\frac{E}{2k_BT}\right)$. $G(\Gamma_D)$ reaches a maximum at $\Gamma_S = \Gamma_D$ where the height of maximum is given by the total tunnel coupling $\Gamma_S + \Gamma_D$ and the temperature $T$. To estimate the change in tunnel coupling $\Gamma_D(d)$ when varying the overlap area between the graphene quantum dot and the graphene drain electrode we performed tight binding calculations of graphene QDs connected to or overlapped with the graphene electrodes (see Fig. S3-S4). We find that $\Gamma_D$ displays oscillations as a function of $d$, possibly due to Fabry-Pérot interferences (19), whose envelope is given by $\Gamma_D(d) = 3d^4$ µeV within the distance range of the simulations, where $d$ is measured in nm (Fig. 5C). Equating the expression for the envelope to $\Gamma_D(d)$ in equation (1), we can fit the data in Fig. 5B where the only free fitting parameter is $\Gamma_S$. We find $\Gamma_S \approx 0.1$ µeV. Our tight binding transport simulations indicate that this small coupling is compatible with narrow source-dot connections having a width of the order of a few nanometers (see section S3).

The capacitive and tunnel coupling data clearly show that displacing the electrodes affects only one side of the junction, as schematically shown in the circuit diagram of the junction (Fig. 5D). The graphene QD is located between the graphene source and drain leads and capacitively coupled to the graphite back gate ($C_G$). During mechanical displacement ($d$) of the graphene leads, the tunnel barrier to the source remains approximately constant, leading to a fixed $\Gamma_S$ and $C_S$. Conversely, $\Gamma_D$ and $C_D$ are modulated mechanically (blue arrow) and show a strong dependence on $d$, with $\Gamma_D$ changing from $10^{-4}$ eV for displacements of 6 nm to $10^{-9}$ eV at zero displacement, which corresponds to a sizeable, five order of magnitude, modulation of the tunnel barrier.

**DISCUSSION**

The uniaxial straining of the monolayer graphene bowtie device has a two effects on the electronic transport measurements. Firstly, it can lead to reduction of out-of-plane height fluctuations (i.e., wrinkles, corrugations), which typically act sources of disorder. Limiting this scattering mechanism has been shown to result in enhanced charge carrier mobility (18), which manifests itself in steeper slopes in the $I$-$V_g$ curves (Fig. 1D and E). Secondly, smoothening out the wrinkles can also account for the formation of a small bilayer graphene overlap region upon closing of the gap (19), as the effective length of the graphene leads is increased.

The subsequent low-temperature transport measurements in the open nanogap regime indicate the presence of a stable, single graphene QD after mechanical breaking of the bowtie. The clean transport features are comparable to those of QDs formed via electrostatic gating in monolayer and bilayer graphene (11, 27). While tearing of monolayer graphene nanoconstrictions is predicted to lead to atomically ordered edges (28), the edges of our device are likely passivated by edge groups during gap opening in air. Our data thus demonstrates that such edge terminations do not adversely impact transport across the gap. We further show that by changing the overlap area between the QD and drain electrode we achieve a high tunability of the tunnel and capacitive coupling, in the former case by over five orders of magnitude. The flexible control over the tunnel barrier strength and symmetry allows us to clearly observe the evolution from a strongly to a weakly coupled graphene QD system.

The formation of graphene QDs has been previously observed in lithographically defined graphene nanoconstrictions (21) and in electroburned nanobowties (29, 30) and is typically attributed to charge localization at graphene edges. In short and narrow constrictions, the regions on which localization occurs can be several tens of nm long (i.e. significantly larger than the dimensions of the constrictions). We suggest that a QD is formed at the edge of the source lead and has a fixed coupling. Our data further indicates that additional QDs form inside the graphene constriction when the gate voltage reaches the CNP. It is unlikely that the additional dots are in parallel to the dominating QD, since these would provide alternative transport pathways and hence lead to non-zero current inside the Coulomb diamonds. Our simulations also exclude a random distribution of energy level characteristic of a chaotic QD as the source of the Coulomb diamond irregularities. We instead conclude that additional QDs are in series with the dominating QD.

Given their strong size dependence on $V_G$, these irregular, non-closing Coulomb diamonds might originate from charge localisation by charge puddles that form around the graphene leads or the graphene QD around the CNP.

In summary, the platform presented here combines the advantages of mechanically controlled two-terminal junctions and three-terminal devices with electrostatic control to probe the electronic transport through graphene nanoconstrictions. Our device design allows for unprecedented tunability of graphene QD transport features and is an alternative approach to the electrostatic control used to tune the coupling strength of QDs in bilayer graphene (11). In particular, the strong asymmetry in tunnel couplings renders this platform attractive in applications such as quantum calorimetry and energy harvesting, where, in the latter case, a direct current can be generated from the exploitation of thermal or voltage fluctuations (15, 16). We anticipate that this experimental platform will also be extended to other 2D materials with the prospect of exploring the low-temperature transport behaviour under electrical and mechanical influence. In particular, it can lend itself to the formation, rupture and controlled overlap of ultra-narrow constriction in superconducting thin films, thereby providing a novel approach to manipulating the Josephson effect in an in-plane device.

## MATERIALS AND METHODS

### Sample Fabrication

The gated graphene break junction devices were prepared based on a previously described protocol (19). Natural graphite flakes were exfoliated onto a PDMS stamp and viewed in an optical microscope in transmission mode (17). Flakes with thicknesses of ~20-30 nm, as confirmed by AFM, were stamped onto $SiO_2$/Si at 120 °C. The same procedure was followed for h-BN flakes of similar thickness. The h-BN/PDMS was positioned such that it covered half of the graphite flake. Graphene was exfoliated onto $SiO_2$/Si and monolayers were identified using Raman spectroscopy and from their contrast in optical microscopy. The graphene flake was picked up using a polycarbonate-coated PDMS stamp (31) and was positioned in the middle of the h-BN flake. The graphene/h-BN/graphite was then picked up and transferred onto pre-prepared phosphor bronze substrates (i.e., coated with polyimide (PI) and polymethylglutarimide (PMGI) and with Au markers already patterned) and released by melting the PC at 180 °C. The sample was then cleaned in anisole at 45 °C for 1 hour and rinsed in IPA and blow dried in $N_2$. The graphene bowties were patterned by exposing the surrounding region (electron dose of 850 $\mu C/cm^2$). The sample was developed in xylene at room temperature for 30 s. The exposed graphene regions were etched by reactive ion etching in an $O_2$ plasma (5 mbar, 20 W, 20 sccm, 5 s). The PMMA was removed by placing the sample in xylene at 80 °C for 10 min. A double layer resist (PMMA A6 495k and A3 950k) was spin coated at 4500 rpm and baked at 180 °C for 2 min. after each layer. Leads and pads were patterned (electron dose of 850 $\mu C/cm^2$) and the sample was developed in xylene at room temperature for 30 s. E-beam evaporation was used to deposit Ti(5 nm)/Au(90 nm). Lift-off was performed in xylene at 80 °C for 30 min.

**Measurement setup**

For the break junction experiments, the junctions are electrically contacted by coating the pads with silver paint. The device was mounted in a three-point bending geometry and operated initially at room temperature in air. Bias voltages of <100 mV and gate voltages between -5 V and 5V were applied to the junction. For low temperature experiments the dipstick was loaded into a cryostat with a base temperature of ~4.2 K, with He exchange gas present in the sample chamber. Bending was performed using a Faulhaber servo motor. Current-voltage measurements were performed using home-built low-noise electronics. For differential conductance measurements a lock-in pre-amplifier (Stanford Research) was used to apply an AC-bias amplitude of 100 µV centered around $V_{SD}$ and with a frequency of 13 Hz.

**Tight binding transport simulations**

We performed simulations of graphene islands attached to graphene electrodes via either narrow constrictions, or via an overlap region. We used a nearest-neighbour tight binding Hamiltonian, where we set the in-sheet hopping energy to $t = 2.7$ eV and modelled the inter-sheet hopping energies similarly to a previous work (32). Both on-site and hopping disorder were in included. Regular hexagonal islands and irregular shapes were investigated. The transport calculations were performed using the code GOLLUM (33).

**Acknowledgments**

**Funding:** S.C. acknowledges a Marie Skłodowska-Curie Individual Fellowship under grant BioGraphING (ID: 798851) and P.G. acknowledges a Marie Skłodowska-Curie Individual Fellowship under grant TherSpinMol (ID: 748642) from the European Union's Horizon 2020 research and innovation programme. This work was supported by the Graphene Flagship (a European Union's Horizon 2020 research and innovation programme under grant agreement No. 649953), the Marie Curie ITN MOLESCO, an ERC advanced grant (Mols@Mols No. 240299) and a Spanish MCIU/AEI/FEDER project (PGC2018-094783).

**Author contributions:** P.G. conceived the idea and with S.C. designed the experiments. S.C. and M.L. prepared the van der Waals heterostructure and carried out the nanofabrication of the samples. M.D.H, S.C. and P.G performed the measurements and the data analysis. A.G-S and J.F. performed the tight binding transport calculations. K.W. and T.T. provided the hBN crystals. P.G, H.v.d.Z and S.C. supervised the project. All authors participated in discussions and co-wrote the paper.

**Competing interests:** The authors declare that they have no competing interests.

**Data and materials availability:** All data needed to evaluate the conclusions in the paper are present in the paper and/or the Supplementary Materials. Additional data related to this paper may be requested from the authors.

**Figures**

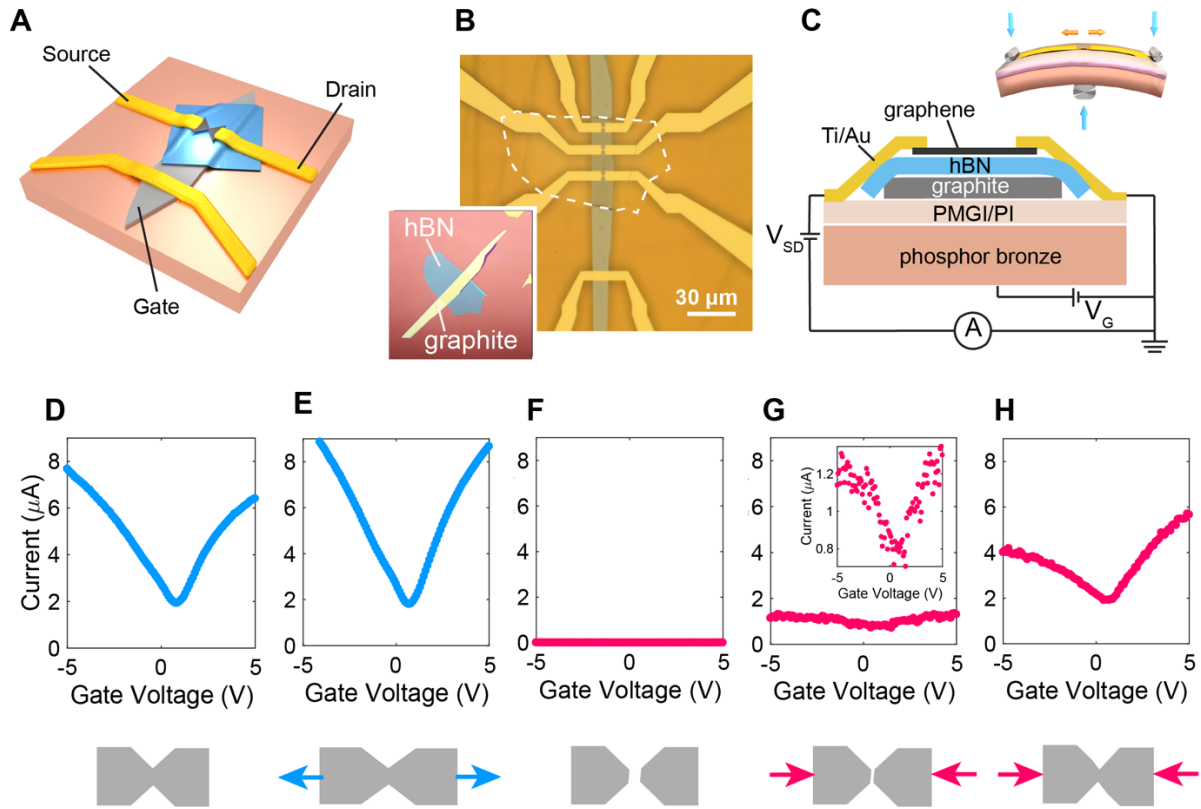

**Fig. 1. Nanofabrication and room-temperature characterization of the graphene MBJ.** **(A)** Schematic of the device layout and **(B)** optical image of three back-gated graphene break junctions. The hBN flake outline is shown by the dotted white line. Bottom inset: van der Waals stack assembled on Si/SiO$_2$ before transfer to the phosphor bronze. **(C)** Side-view of the device composed of a graphene/hBN/graphite stack on a polymer-coated flexible metal substrate. Inset: 3-point bending experiment. The evolution of the Dirac curve for **(D)** unstrained, **(E)** strained, **(F)** broken and **(G-H)** remaking after rupture. A schematic of the junction configuration is shown below each plot. All plots were acquired with a bias voltage of 100 mV.

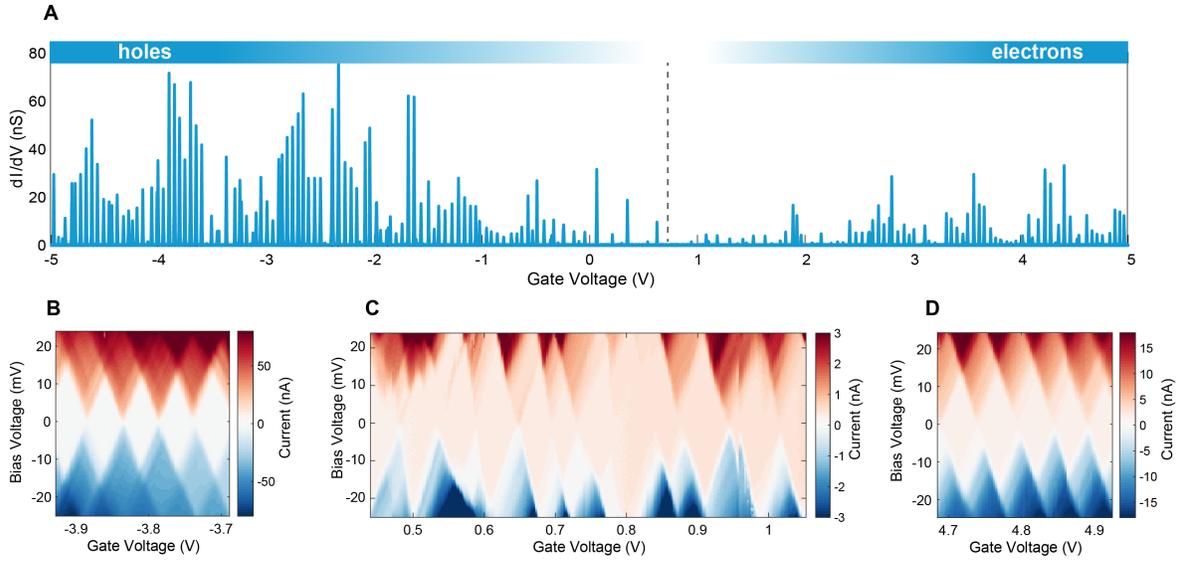

**Fig. 2. Coulomb blockade at 4 K. (A)** Conductance peaks measured over the whole gate range (±5 V) showing suppressed conductance around 0 V-2 V and sharp resonances in the electron and hole regimes. The vertical grey dotted line indicates the position of the CNP before breaking the junction. Stability diagrams of the current as a function of $V_{SD}$ (± 25 mV) and $V_G$ measured at different gate regions: **(B)** Hole regime, **(C)** at the charge neutrality point and **(D)** electron regime.

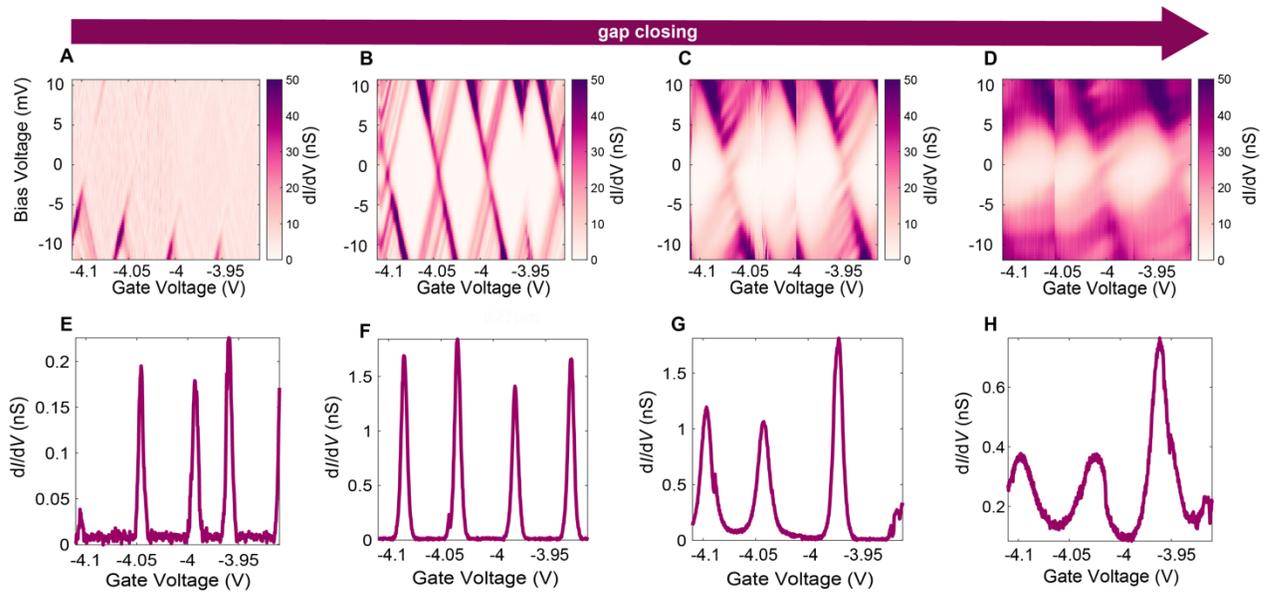

**Fig. 3. Mechanical tuning of tunnel couplings. (A-D)** Differential conductance as a function of bias and gate voltage during mechanical closing of the junction in the hole regime. **(E-H)** Corresponding conductance peaks measured with a lock-in pre-amplifier with an AC-bias amplitude of 100 µV.

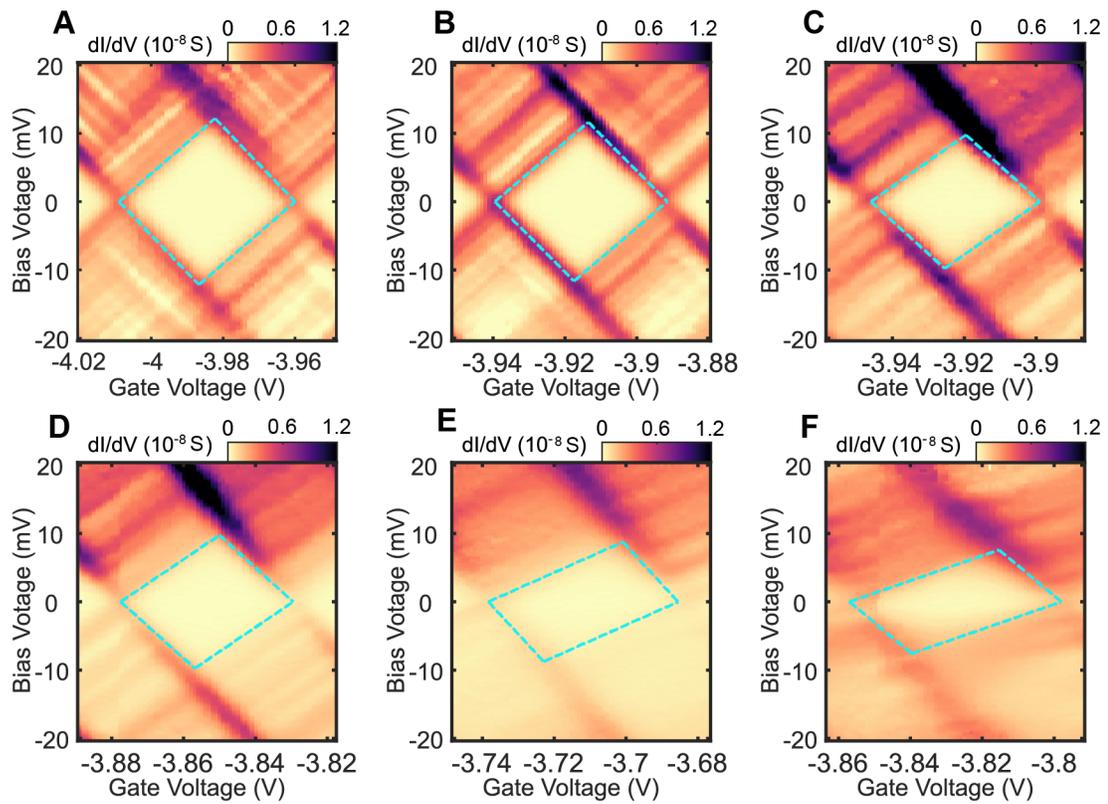

**Fig. 4. Mechanical tuning of capacitive couplings.** Differential conductance map of a selected Coulomb diamond for in-plane displacements (i.e., graphene overlap lengths) of **(A)** 0.8 nm, **(B)** 1 nm, **(C)** 1.3 nm, **(D)** 2.1 nm, **(E)** 2.7 nm and **(F)** 3.4 nm.

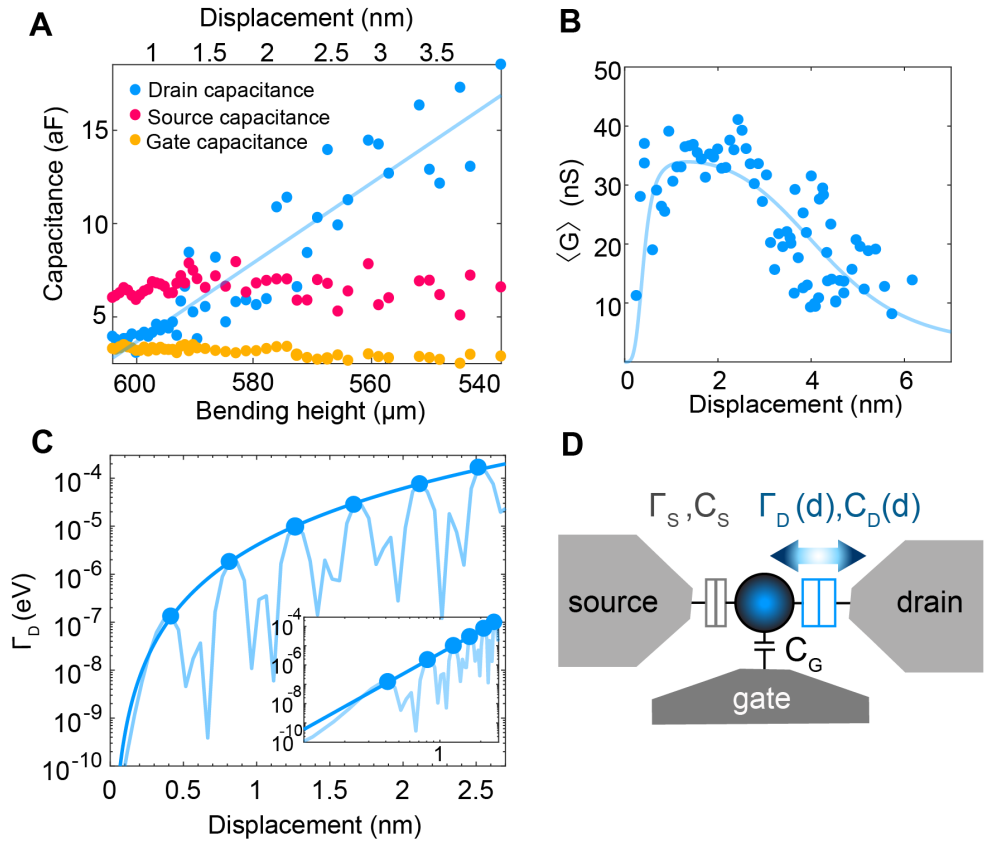

**Fig. 5. Quantification of the capacitive and tunnel coupling asymmetries.** **(A)** Capacitances as a function of bending height (bottom *x*-axis) and in-plane displacement (i.e., overlap length, top *x*-axis) of the gate (yellow), source (magenta) and drain (blue), extracted from the slopes of the diamonds in Fig. 4. **(B)** Mean height of the conductance peaks in the gate voltage window between -4.1 V and -3.5 V, as a function of displacement (circles). The solid line indicates the fit to the data using a Landauer approach at a fixed temperature 4.2 K, a calculated source tunnel coupling, $\Gamma_S$, of 0.1 µeV, a graphene layer separation of 0.335 nm and an overlap region width of 160 nm. **(C)** Drain tunnel coupling as a function of graphene overlap displacement. The inset shows the same coupling on a logarithmic scale. **(D)** Schematic illustration of the mechanical quantum dot model: the QD is located on the source-side of the graphene junction. The drain edge moves closer (further) to it during closing (opening) giving rise to the variation in the drain tunnel and capacitive couplings.